\documentclass[letterpaper,twocolumn]{jpsj3}
\usepackage{amsfonts}
\usepackage[T1]{fontenc}
\usepackage{txfonts}
\usepackage{amsmath}
\usepackage{amssymb}
\usepackage{graphicx}
\usepackage{color}
\usepackage{tensor}
\usepackage{bm}
\usepackage{comment}
\usepackage{mathrsfs}
\usepackage{multicol}
\usepackage[version=4]{mhchem}
\usepackage[svgnames]{xcolor}

\inst{$^1$Division of Natural and Environmental Sciences, The Open University of Japan, Chiba, 261-8586, Japan \\
$^2$Department of Physics and Electronics, Osaka Metropolitan University, 1-1 Naka-ku, Sakai, Osaka 599-8531, Japan  \\
$^3$Quantum Research Center for Chirality, Institute for Molecular Science, Okazaki, Aichi 444-8585, Japan\\
$^4$ Department of Basic Science, The University of Tokyo, 3-8-1 Komaba, Tokyo 153-8902, Japan} 
\abst{We explicitly derive the wavefunctions of chiral phonons propagating along the helical axis in chiral crystals and clarify the characteristics of electron-phonon interactions in chiral helical crystals. In particular, we elucidate how the conservation of not only the crystal momentum (CM) but also the crystal angular momentum (CAM) manifests in the interaction vertex. This formulation provides a microscopic framework for describing physical processes involving chiral phonons. Furthermore, we construct a phononic analogue of Zilch, a known measure of chirality carried by light, and discuss its relationship with phonon angular momentum.}

\begin{document}

\title{Electron-Chiral Phonon Coupling, Crystal Angular Momentum, and Phonon Chirality}
\author{Tomomi Tateishi$^1$\thanks{%
t.tateishi@ouj.ac.jp}, Akihito Kato$^{2}$ and Jun-ichiro Kishine$^{1,3,4}$}
\maketitle

Recently, research on chiral phonons---the quanta of lattice wave in which the rotational motion of atoms propagates with a non-zero wave number along the rotational axis---has been attracting considerable attention.
Chiral phonons are clearly realized along the helical axis of a chiral crystal.  Though research on phonons in chiral crystals dates back more than half a century,~\cite{Pine1969,Teuchert1974} recent works link  phonons to the angular momentum emphasizing their rotational nature.~\cite{Vonsovskii1962,Zhang2014,Zhang2015,Zhang2022,Tsunetsugu2023} Two key aspects drive this renewed
interest. First, phonon dispersion splits depending on left- or right-handedness of crystal.~\cite{Pine1969,Teuchert1974,Kishine2020,Tsunetsugu2023} 
Second, chiral phonons possess two types of angular momenta: mechanical angular momentum (MAM) associated with atomic orbital motion around an equilibrium position,~%
\cite{Vonsovskii1962} which is non-conserved due to broken continuous rotational symmetry of a crystal, and conserved crystal angular momentum (CAM) arising from discrete rotational symmetry.\cite{Zhang2014,Zhang2015,Hamada-Minamitani-Murakami2018,Hamada2020,Zhang2022,Tsunetsugu2023} The concept of CAM was early pointed out for electrons propagating in helical crystals.\cite{Bozovic1984}. The conservation and interconversion of CAM among phonons, electrons, and light in chiral crystals was recently experimentally confirmed by Raman selection rules.\cite{Ishito2022,Ishito2023,oishi2024}

Unlike conventional phonons, a hallmark of chiral phonons is the mixing of longitudinal and transverse modes.\cite{Hamada-Minamitani-Murakami2018,Hamada2020,Ishito2022,Zhang2022}
Two of the present authors recently derived phonon wavefunctions in chiral helical crystals using group representation theory.~\cite{Kato2023} In this letter, we apply this to the description of electron-phonon interactions. In contrast to conventional cases,  where electrons couple only to longitudinal phonons, chiral helical crystals permit coupling of electrons to left- and right-handed chiral rotational modes, and consequently the electron-phonon interaction vertex carries CAM. We also discuss phonon analogue of Lipkin's Zilch,~\cite{Lipkin1964}  which is a time-even pseudoscalar\cite{Barron1982} that serves as a chirality measure for circularly polarized light.\cite{Tang-Cohen2010}
 
\begin{figure}[ptb]
\begin{center}
\includegraphics[width=6cm]{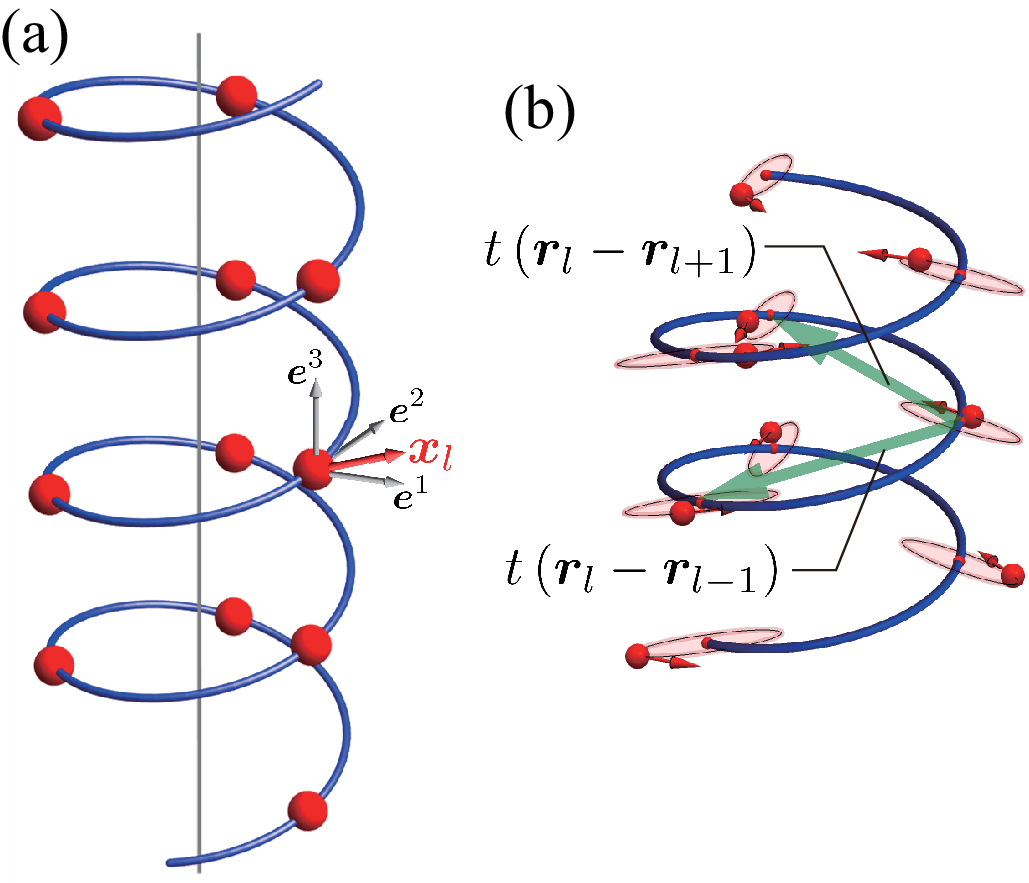} 
\end{center}
\caption{(Color online).(a) Helical crystal with symmetry of the line group $L3_{1}$. (b)
Schematic illustration of atomic motion (red spheres) and electron hopping
paths (green arrows) for electrons localized at atoms.}
\label{helical_crystal}
\end{figure}

With a chiral crystal \ce{Te} (tellurium) in mind, as
shown in Fig.~\ref{helical_crystal}(a), we consider a model crystal
belonging to the line group $L3_{1}  =\{\mathcal{\hat{R}}^{p}\mid
p=0,1,\ldots,3N-1\}$, where $N$ is the total number of the unit cell, and analyze phonons propagating along the helical axis. Though the following method for deriving the phonon wavefunction was briefly outlined in Ref.~[\citen{Kato2023}], here we present the essential aspects required to establish a consistent framework that incorporates the electronic wavefunction and describes the electron-phonon interaction. 
The
threefold screw operator, $\mathcal{\hat{R}}=(\mathcal{\hat{C}}_{3}\,|1/3)$,
acts on a position vector $\bm{r}$ as $\mathcal{\hat{R}}\bm{r} = \mathcal{%
\hat{C}}_3\bm{r} + (1/3) c\bm{e}^3$, where $\mathcal{\hat{C}}_3$ is the
threefold rotation operator and $c$ is the lattice constant along the screw
axis with $\bm{e}^1, \bm{e}^2, \bm{e}^3$ being the Cartesian basis vectors.
The irreducible representations(irreps.) of $L3_1$ are classified by $%
n=0,1,\cdots N-1$ with the crystal momentum (CM), $q_{n}\equiv\frac{2\pi n}{cN%
}$, originated from the discrete translation symmetry and the CAM, $m=0,\pm1$%
, from the threefold rotation by an angle $\alpha\equiv2\pi/3$. Then, the
character of the irreps. for $\mathcal{\hat{R}}^p$ is given by $%
\chi_{n,m}(p)=\exp[ i( q_{n}\frac{c}{3}+\alpha m) p]$.~\cite%
{Bozovic1978,Kato2023}

Introducing the chiral basis, $\bm{e}^{\pm}\equiv \frac{1}{\sqrt{2}}(\bm{e}%
^{1} \mp  i\bm{e}^{2})$ and $\bm{e}^{0}\equiv\bm{e}^{3}$, the equilibrium
atomic position is represented as $\bm{R}_{l} = \frac{\rho}{\sqrt{2}}
e^{i(l-1)\alpha} \bm{e}^{+} + \frac{\rho}{\sqrt{2}}  e^{-i(l-1)\alpha} \bm{e}%
^{-} +\frac{c}{3}(l-1) \bm{e}^0$ ($l=1, 2, \ldots,  3N-1$), where $\rho$
denotes the radius of the helix and $l$ is a sequential label that
designates atoms along the helix one by one from the bottom. The displaced
atomic position is expressed as $\bm{r}_{l}\left(t\right) =\bm{R}_{l}+\bm{u}%
_{l}\left( t\right) $, with $\bm{u}_{l}\left( t\right) 
=\sum_{s=\pm,0}u_{l}^{s}\left( t\right) \bm{e}^{s}$, using chiral basis.

To describe symmetry operations on the entire system, we introduce a $9N$%
-dimensional vector, denoted as $\bm{X}(t) = \sum_{l=1}^{3N} \sum_{s=\pm,0}
u_{l}^{s}(t) \bm{e}_{l}^{s}$. Here, we define the basis assigned to the $l$%
-th site as $\bm{e}_{l}^{s} = \bm{e}_{l} \otimes \bm{e}^{s}$, where $\bm{e}%
_{l}$ is a unit vector in the $3N$-dimensional space, given by $\bm{e}_{l} =
(0,\ldots,0,1,0,\ldots,0)^{T}$, with the only nonzero element being $1$ at
the $l$-th position. The superscript $T$ denotes the transpose.

The chiral basis $\bm{e}_{l}^{s}$ are eigenvectors of $\mathcal{\hat{C}}_{3}$%
, satisfying $\mathcal{\hat{C}}_{3} \bm{e}_{l}^{s} = e^{i m_s \alpha} \bm{e}%
_{l}^{s}$, where $s = \pm, 0$ and $(m_{+}, m_{-}, m_{0}) = (1, -1, 0)$.
Hence, $\mathcal{\hat{R}}$ acts on $\bm{e}_{l}^{s}$ as $\mathcal{\hat{R}}%
\bm{e}_{l}^{s}=e^{im_{s}\alpha}\bm{e}_{l+1}^{s}$. This allows us to
construct the projection operator, $\mathcal{\hat{P}}_{n,m}=\frac{1}{\sqrt{3N}}\sum_{l=1}^{3N} \chi_{n,m}^{\ast}(l-1)\mathcal{\hat{R}}^{l-1}$.
The symmetry-adapted basis are then given by $\bm{e}_{n,m}^{s}\equiv\mathcal{%
\hat{P}}_{n,m} \bm{e}_{l=1}^{s}$, with $\bm{e}_{l=1}^{s}$ being a trial
vector.

Using this symmetry-adapted bases, we have $\bm{X}(t) = \sum_{n=0}^{N-1}
\sum_{m=\pm1,0}  \sum_{s=\pm, 0} w_{nm}^{s}(t)\bm{e}_{n,m}^{s}$,  where we
introduce the time-dependent normal coordinates, $w_{nm}^{s}\left(t\right)$,
expressed as 
\begin{equation}
u_{l}^{s}(t) = \frac{1}{\sqrt{3N}} \sum_{n=0}^{N-1} \sum_{m=\pm1, 0}
e^{-i(l-1) [ q_{n} \frac{c}{3}+\alpha(m-m_{s})]} w_{nm}^{s}(t).  \label{uls}
\end{equation}
The appearance of $m_{s}=0, \pm 1$ reflects the three-dimensional nature of
lattice vibrations. This transformation is complemented by the orthogonality
relation and the following relation, $\sum_{n=0}^{N-1}\sum_{m=%
\pm1,0}e^{i(l-l^{\prime})\left( \frac{q_{_{n}}c}{3}+\alpha m\right) }
=3N\delta_{l,l^{\prime}}$.

\begin{figure}[ptb]
\centering
\includegraphics[width=8.5cm]{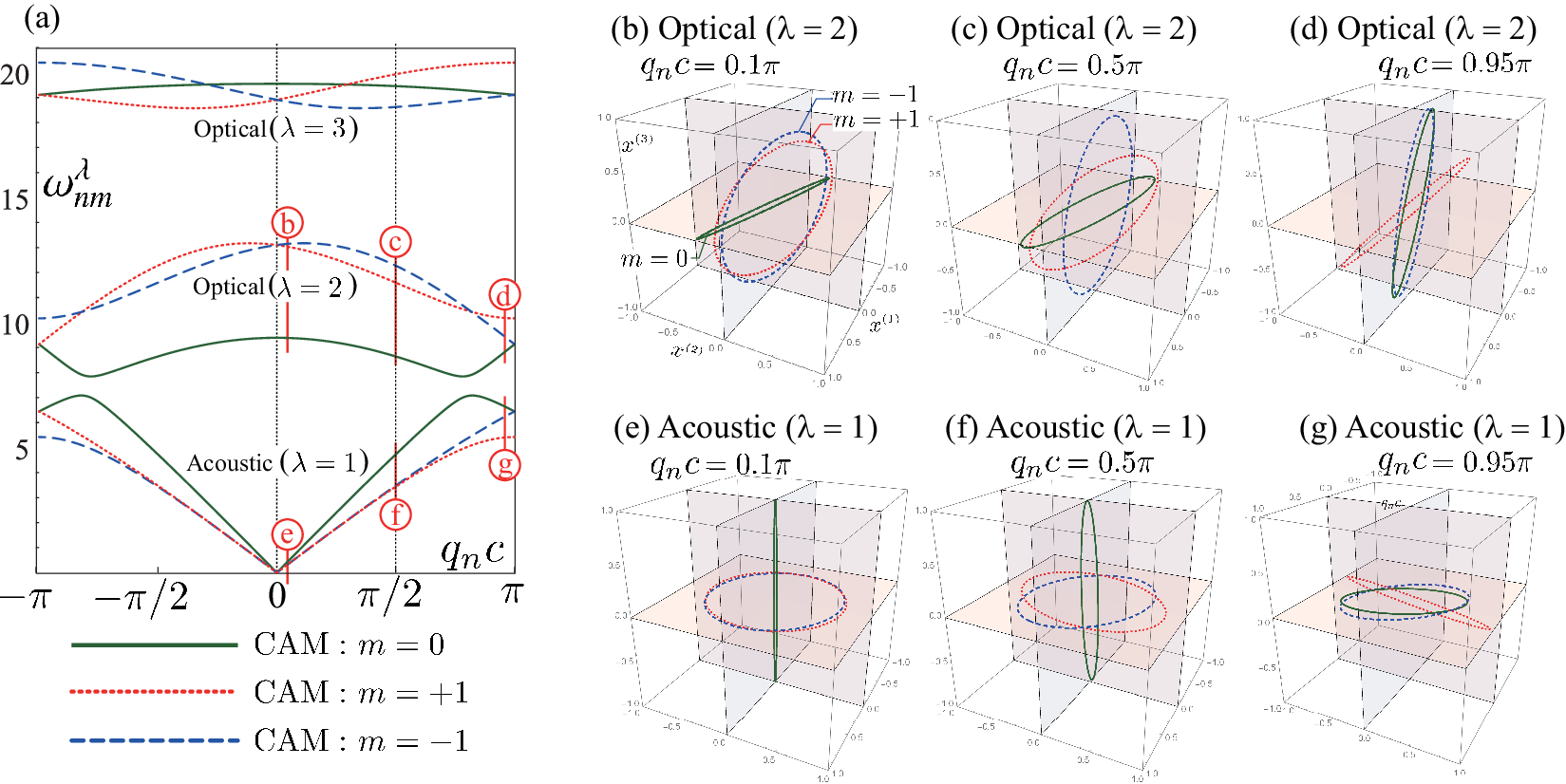}  
\caption{(Color online).(a) Band dispersion of lattice vibrations for a chiral crystal
belonging to the line group $L3_{1}$. The force constant matrix elements are
set as $\tilde{\Omega}_{11}=\tilde{\Omega}_{22}=78.4$, $\tilde{\Omega}
_{33}=98.0$, $\tilde{\Omega}_{12}=49.0$, $\tilde{\Omega}_{13}=\tilde{\Omega}
_{23}=29.4$. All units are in [meV]. (b)-(g): Real-space trajectories of
atomic motion at site $l=1$ for crystal momenta $q_nc=0.1\protect\pi$, $0.5%
\protect\pi$, and $0.95\protect\pi$. Trajectories corresponding to the first
optical branch ($\protect\lambda=2$, (b)-(d)) and the acoustic branch ($%
\protect\lambda=1$, (e)-(g)) are shown. The corresponding locations on the
dispersion curves are indicated by circled labels in (a). }
\label{Band_dispersion}
\end{figure}

In the harmonic approximation, the potential energy of the lattice vibration
is given by $\Phi=\frac{M}{2}\sum_{l=1}^{3N} 
\sum_{s^{\prime},s=\pm,0}(u_{l+1}^{s^{\prime}}-u_{l}^{s^{\prime}})^{\ast}
\Omega_{s^{\prime}s}^{(l)}(u_{l+1}^{s}-u_{l}^{s})$, where $M$ is the ionic
mass and $\Omega^{(l)}_{s^{\prime }s}$ is the force constant matrix element
in the chiral coordinates with the following elements: $\Omega_{++}=
\Omega_{--}=\frac{1}{2}(\tilde{\Omega}_{11}+\tilde{\Omega}_{22})$, $
\Omega_{+-}= \Omega_{-+}^{\ast }=\frac{1}{2}(\tilde{\Omega}_{11}-\tilde{
\Omega}_{22}+2i\tilde{\Omega}_{12})$, $\Omega_{+0}=\Omega_{0+}^{\ast}=\frac{%
1 }{\sqrt{2}}(\tilde{\Omega}_{13}+i\tilde{\Omega}_{23})$, $\Omega_{-0}=
\Omega_{0-}^{\ast}= \frac{1}{\sqrt{2}}(\tilde{\Omega}_{13}-i\tilde{\Omega}
_{23})$, and $\Omega_{00}=\tilde{\Omega }_{33}$, with $\tilde{\Omega}
^{(l)}_{ij}$ representing that in the Cartesian coordinates which must be a
positive definite matrix. For simplicity, we have omitted the index $l$.
Substituting Eq.~\eqref{uls} into the potential energy, we obtain $\Phi= 
\frac{M}{2}\sum_{n=0}^{N-1}\sum_{m=\pm1,0}\bm{w}_{nm}^{\dagger }\mathbf{A}
_{nm}\bm{w}_{nm}$, with $\bm{w}_{nm}\equiv 
(w_{nm}^{(+)},w_{nm}^{(-)},w_{nm}^{(0)})^{T}$. The dynamical matrix elements
are $(\mathbf{A}_{nm})_{s^{\prime}s}\equiv(d_{nm}^{s^{\prime}})^{\ast}%
\Omega_{s^{\prime}s}^{(1)}d_{nm}^{s}$,  where we defined the phase
difference factor, $d_{nm}^{s}\equiv e^{-i[q_{n}\frac{c}{3}%
+\alpha(m-m_{s})]}-1 \label{dsnm}$. Solving the eigenvalue equation, $%
\mathbf{A}_{nm}\bm{v}_{nm}^{\lambda}=( \omega_{nm}^{\lambda})^{2}\bm{v}%
_{nm}^{\lambda}$, for $\lambda=1,2,3$, we obtain the phonon dispersion, which is depicted in Fig.~\ref{Band_dispersion}(a).
For the zero CM ($q_{n}=0$), \(d_{nm}^s\) vanishes at $m = m_s$, yielding three zero-frequency eigenvalues with uniform displacements. These correspond to the acoustic mode, labeled $ \lambda = 1 $, while the nonzero-frequency optical modes are assigned $\lambda = 2,3$.
The $s$-th component of the eigenvector, $%
v_{n,m}^{\lambda (s)}$, satisfies $(v_{n,m}^{\lambda(s)})^{\ast}=v_{-n,-m}^{%
\lambda(-s)}$, which is a consequence of the time reversal symmetry. 
The decomposition, $\bm{w}_{nm}\equiv\sum_{\lambda}c_{nm}^{\lambda}\bm{v}
_{nm}^{\lambda}, \label{wtildelam}$ leads to the diagonalized potential, $
\Phi=\frac{M}{2}\sum_{n=0}^{N-1}\sum_{m=\pm1,0}\sum_{{\lambda=1,2,3}
}|c_{nm}^{\lambda}|^{2}(\omega_{nm}^{\lambda})^{2}$, where the coefficients $
c_{nm}^{\lambda}$ play a role of the normal coordinates.

Since the explicit form of the wave function has been obtained, it becomes
possible to depict the trajectory of atomic motion in real space. Classical
trajectories are obtained from the equation of motion for $%
c_{n,m}^{\lambda}(t)$. We pick up the position $l=1$, and select a
particular frequency mode specified by $n, m$, and $\lambda$. This leads to
the equation $u_{l}^{(s)}\left( t\right) =\frac{1}{\sqrt{3N}}(\eta_{n,m}^{
\lambda1}e^{i\omega_{n,m}^{\lambda}t} 
+\eta_{n,m}^{\lambda2}e^{-i\omega_{n,m}^{\lambda}t})v_{n,m}^{\lambda(s)}$.
where $\eta_{n,m}^{\lambda1}$ and $\eta_{n,m}^{\lambda2}$ are the
integration constants. In Figs. \ref{Band_dispersion} (b)-(g), we present
the real-space trajectories. We observe that as the CM increases while
keeping the CAM fixed, the orbital eccentricity of the atomic motion deforms
and the orbital plane rotates. For example, the acoustic mode with $m=1$ exhibits a
horizontal transverse orbit near the $\Gamma$ point [(e)], but transforms
into a longitudinal orbit near the Brillouin zone boundary [(g)]. This
situation indicates that for chiral phonons, contrary to the conventional
understanding, the longitudinal and transverse components mix at non-zero
wave numbers, and consequently the chiral transverse modes contribute to the
electron-phonon coupling, as discussed below.

Introducing the canonical momentum, $p_{n,m}^{\lambda}=M\dot{c}%
_{n,m}^{\lambda}$, the classical Hamiltonian becomes $\mathscr{H}_{\text{lat}%
}= {\textstyle\sum\limits_{n,m,\lambda}} \left[ \frac{1}{2M}%
|p_{n,m}^{\lambda}|^{2}+\frac{M}{2}(\omega_{n,m}^{
\lambda})^{2}|c_{n,m}^{\lambda}|^{2}\right]$.
We promote the canonical
variables $p_{n,m}^{\lambda}$ and $ c_{n,m}^{\lambda} $ to operators
satisfying the fundamental commutation relations: $\lbrack\hat{c}%
_{n,m}^{\lambda},\hat{p}_{n^{\prime},m^{\prime}}^{
\lambda^{\prime}}]=i\hbar\,\delta_{n,n^{\prime}}\delta_{m,m^{\prime}}
\delta_{\lambda,\lambda^{\prime}}$, and construct the phonon annihilation
operators, defined by $\hat{a}_{n,m}^{\lambda}=(\frac{\omega_{n,m}^{\lambda} 
}{2\hbar M})^{\frac{1}{2}}(M\hat{c}_{n,m}^{\lambda}+\frac{i}{
\omega_{n,m}^{\lambda}}\hat{p}_{-n,-m}^{\lambda})$, which satisfy $[\hat{a}
_{n,m}^{\lambda},\hat{a}_{n^{\prime},m^{\prime}}^{\lambda^{\prime}\dagger}]=
\delta_{n,n^{\prime}}\delta_{m,m^{\prime}}\delta_{\lambda,\lambda^{\prime}}$%
.  We then express the quantized phonon field operator as 
\begin{equation}
\hat{\bm{w}}_{nm}(t)=\sum_{\lambda=1,2,3}\left( \frac{\hbar} {%
2M\omega_{nm}^{\lambda}}\right) ^{\frac{1}{2}}(\hat{a}_{n,m}^{\lambda }(t) +%
\hat{a}_{-n,-m}^{\lambda\dagger}(t))\bm{v}_{nm}^{\lambda},  \label{quantwnm}
\end{equation}
where $\hat{\bm{w}}_{nm}(t)$ denotes the quantized version of $\bm{w}_{nm}(t)
$.

To describe the electron--phonon interaction, we adopt the modified
tight-binding approximation~\cite{Friedel1970}, which allows us to respect
the crystal symmetry of electron hopping along the helix, as depicted in
Fig.~\ref{helical_crystal}(b). We start with a modified tight-binding
Hamiltonian, 
\begin{equation}
\mathscr{H}_{\text{el}}=-\sum_{\langle
l,l^{\prime}\rangle}\sum_{\sigma}t\left(\bm{r}_{l}-\bm{r}_{l^{\prime}}
\right) \hat {c}_{l,\sigma}^{\dagger}\hat{c}_{l^{\prime},\sigma},
\label{MTBH}
\end{equation}
where $\hat{c}_{l,\sigma}^{\dagger}$ and $\hat{c}_{l,\sigma}$ are the
creation and annihilation operators of electrons, with the atomic orbitals
centered around the displaced ionic positions, $\bm{r}_{l} =\bm{R} _{l}+%
\bm{u} _{l}(t)$. The spin index is represented by $\sigma
=\uparrow,\downarrow$, and the hopping is assumed to be restricted to the
nearest-neighbor sites $ \langle l,l^{\prime}\rangle$. Expanding the overlap
integral, $t( \bm{r}_{l}  - \bm{r}_{l^{\prime}} )$, to first order in $\bm{u}%
_{l}(t)-\bm{u} _{l^{\prime}}(t)$ yields 
\begin{equation}
t\left( \bm{r}_{l}-\bm{r}_{l^{\prime}}\right) =t_{0}+\left. \frac{\partial
t\left( \bm{r}\right) }{\partial\bm{r} } \right\vert _{\bm{r}=\bm{R}_{l}- %
\bm{R}_{l^{\prime}}}\cdot \left[ \bm{u}_{l}(t)-\bm{u}_{l^{\prime}}(t)\right]
,
\end{equation}
where $t_{0}\equiv t\left( \bm{R}_{l}-\bm{R}_{l^{\prime}}\right) $.

Following the idea of Ref.~[\citen{Friedel1970}], we assume the symmetry of
the wavefunction with respect to the helical lattice implies that the
gradient of the overlap integral aligns with the helical bonds.
The electron orbitals are aligned according to the operations of the line group \( L3_{1} \) along the helix. Therefore, symmetry constrains the electron hopping paths, and this approximation is expected to be justified regardless of the orbital type (e.g., \( s \), \( p \), \( d \), etc.).\cite{Friedel1970}.
Since the Fourier transform of $\hat{c}_{l ,\sigma }$ is given by $\hat{c}_{l,\sigma }
=\frac{1}{\sqrt{3N}}\sum_{n^{\prime },m^{\prime }}  e^{-i(l-1)\left(
k_{n^{\prime }}\frac{c}{3}+\alpha m^{\prime }\right) }\hat{c}_{n^{\prime
},m^{\prime },\sigma }$,~\cite{Bozovic1984} the electron--chiral phonon
coupling Hamiltonian is obtained as 
\begin{align}
\mathscr{H}_{\text{el}-\chi\text{ph}} & =\zeta t_{0} \sum_{\langle
l,l^{\prime }\rangle}\sum_{\sigma } \left( \frac{\bm{R}_{l}-\bm{R}%
_{l^{\prime}}} { \left\vert \bm{R}_{l}-\bm{R}_{l^{\prime}}\right\vert}%
\right)\cdot \left[\hat{ \bm{u}}_{l}(t)-\hat{\bm{u}}_{l^{\prime}}(t)\right] 
\hat{c} _{l,\sigma}^{\dagger}\hat{c}_{l^{\prime},\sigma}  \notag \\
& = \sum_{n,n^{\prime},m,m^{\prime}}\sum_{\lambda, \sigma} \Gamma_{n
n^{\prime} m m^{\prime}}^{\lambda} (\hat{a}_{n,m}^{\lambda}+\hat{a}
_{-n,-m}^{\lambda\dagger})\hat{c}_{n+n^{\prime},m+m^{\prime},\sigma}^{
\dagger}\hat{c}_{n^{\prime},m^{\prime},\sigma},
\end{align}
where $\zeta$ represents the Slater coefficient describing the exponential
decay $e^{-\zeta r}$ of the electronic Wannier function,~\cite{Friedel1970}
and $\hat{\bm{u}}_{l}(t)$ stands for the quantized version of $\bm{u}_{l}(t)$%
. The vertex function is explicitly written as 
\begin{align}
& \Gamma_{nn^{\prime }mm^{\prime }}^{\lambda} = \left( \frac{\hbar }{
6MN\omega _{nm}^{\lambda }}\right)^{\frac{1}{2}} \frac{\zeta t_{0}}{\sqrt{
3\rho^{2}+c^{2}/9}}  \notag \\
& \times \left[ \sqrt{6} \rho \left(\gamma_{nn^{\prime }mm^{\prime }}^{(+)}
v_{nm}^{\lambda (+)} - \gamma_{nn^{\prime }mm^{\prime }}^{(-)}
v_{nm}^{\lambda (-)} \right)+ \frac{2i}{3}c \gamma_{nn^{\prime }mm^{\prime
}}^{(0)} v_{nm}^{\lambda (0)} \right],  \label{eq:vertex}
\end{align}
where the coherence factor, $\gamma_{nn^{\prime }mm^{\prime
}}^{(s)}=g_{n^{\prime },m^{\prime}+m_s/2}-g_{n+n^{\prime },m+m^{\prime
}-m_s/2}$ with $g_{n,m}=\sin(q_nc/3+\alpha m)$, represents interference
effect reflecting chiral nature of both propagating electrons and phonons.
Equation~\eqref{eq:vertex} is the central result of this study. Note that
electron hopping along the sites on the helix induces a direct coupling with
transverse phonons; this is the reason for the emergence of $v_{nm}^{\lambda
(\pm)}$. We note that in the case of $\rho=0$, while keeping $c$ nonzero,
the second term in the parentheses of Eq.~\eqref{eq:vertex} reproduces the
same result as given in Ref.~[\citen{Friedel1970}], while the first term
which contains $\rho$ are specific to the helical crystal. It is to be noted
that taking a naive continuum limit, $c\to 0$, is meaningless in helical
crystals. We also note that the vertex $\Gamma_{n n^{\prime} m
m^{\prime}}^{\lambda}$ satisfies $\Gamma_{-n, -n^{\prime}, -m,
-m^{\prime}}^{\lambda}=(\Gamma_{n n^{\prime} m m^{\prime}}^{\lambda})^\ast$,
which is a consequence of time-reversal symmetry. The essential point is
that at the electron-phonon interaction vertex, not only the CM but also the CAM must be conserved.

The behavior of the vertex function is briefly discussed below. For instance, when considering thermal phonon excitation at temperatures much lower than the Debye temperature, the phonon wave vector satisfies \( |q_n| \ll \pi/c \). In this case, since the vertex has a square root of $\omega_{nm}^{\lambda}$ in the denominator, for the acoustic mode($\lambda=1$), although $ \gamma_{nn'mm'}^{s}$ approaches zero for terms that satisfy $m=m_{s}$, the corresponding component of the eigenvector, $v_{nm}^{\lambda(s)}$, takes nonzero values. When $m\neq m_{s}$, although $\gamma_{nn'mm'}^{s}$ takes nonzero values, $v_{nm}^{\lambda(s)}$ approaches zero. In either case, while the whole expression becomes indeterminate, the value converges. 
On the other hand, for the optical modes ($\lambda=2,3$), although the contribution from the square root of $\omega_{nm}^{\lambda}$ in the denominator is small, the combination of when $\gamma_{nn'mm'}^{s}$ and $v_{nm}^{\lambda(s)}$ become zero or nonzero is the opposite of the case for $\lambda=1$.
Therefore, the total contribution becomes slightly larger than that for $\lambda=1$.
In reality, we need to incorporate information from the electron band branch and the phonon CAM. The behavior of this function is further constrained depending on the information of the Fermi level.

\begin{figure}[ptb]
\begin{center}
\includegraphics[width=8.5cm]{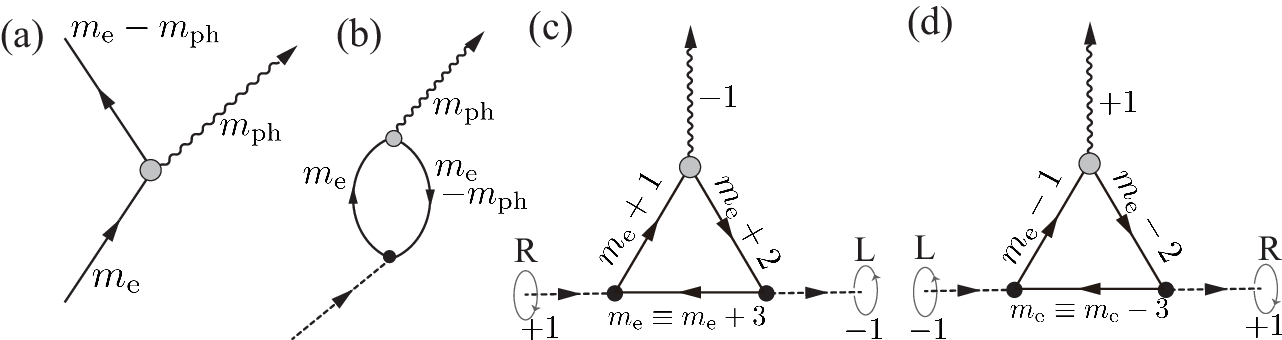} 
\end{center}
\caption{(Color online).(a) Electron-chiral phonon interaction. (b) External-field-induced
electron-hole excitation that recombines to emit a chiral phonon with
nonzero CAM. (c) and (d) Stokes and anti-Stokes Raman processes,
respectively, mediated by a chiral phonon. The solid, dashed, and wavy lines
respectively represent an electron, photon, and phonon propagators.}
\label{Feynman}
\end{figure}

The details of the physical processes induced by the electron--chiral phonon
interaction of Eq.~\eqref{eq:vertex} will be left for future studies;
Instead, we demonstrate a few relevant processes using Feynman diagrams
[Figs.~\ref{Feynman}(a)-(d)]. Fig.~\ref{Feynman}(a) illustrates the
elementary process of electron-phonon interaction, highlighting the
conservation law of CAM at the vertex.

Fig.~\ref{Feynman}(b) illustrates the process in which an electron-hole pair
is excited in the electronic subsystem by an external field (e.g., an
electric field) and, upon recombination, emits a chiral phonon with non-zero
CAM. This process indicates, for instance, the possibility that an electric
current induces phonon angular momentum, which, if transferred to the
electron spin, could lead to long-range spin polarization. The relationship
between this process and the chirality-induced spin selectivity over
macroscopic scales in inorganic chiral crystals~\cite{inui2020,shiota2021}
is an intriguing issue and will be addressed in future studies.

Furthermore, Figs.~\ref{Feynman}(c) and (d) respectively show the Stokes and
anti-Stokes Raman scattering processes mediated by chiral phonons. These
diagrams illustrate the kinetic constraint on the loop process in which an
electron receives CAM first from a photon and then from a phonon before
returning to its initial state. In the case of a threefold helical system,
it is crucial that CAM conservation holds modulo 3, i.e., $m_{\mathrm{e}%
}\equiv m_{\mathrm{e}}\pm 3$.~\cite{Tatsumi2018} Corresponding to the
experiment in Ref.~[\citen{Ishito2022,Ishito2023,oishi2024}], as shown in
Fig.~\ref{Feynman}(c), when the incident light is right-handed circularly
polarized (photon spin $\sigma_i = +1$) and is emitted as left-handed
circularly polarized light ($\sigma_f = -1$), the emitted phonon's CAM is
restricted to $m_{\mathrm{ph}} = -1$, as long as umklapp scattering is
negligible. Fig.~\ref{Feynman}(d) shows the case where the combination of
incident and emitted circularly polarized light is reversed.

Finally, we discuss the MAM and Zilch of chiral phonons. The phonon MAM in
the $z$-direction is given by $J_{z}=M\sum_{l=1}^{3N}(x_{l}^{(1)}\dot{x}%
_{l}^{(2)}-x_{l}^{(2)}\dot{x}_{l}^{(1)})$. Using the explicit form of phonon
wavefunctions, the quantized form of the MAM is obtained, and its thermal
average is given by 
\begin{align}
& \langle\hat{J}^{z}\rangle_\mathrm{eq} \equiv \sum_{n=-\frac{N-1}{2}}^{%
\frac{N-1}{2}}\sum_{m=\pm1,0}\sum_{\lambda=1,2,3}\langle\hat{J}
_{nm\lambda}^{z}\rangle_\mathrm{eq}, \\
& \langle\hat{J}_{nm\lambda}^{z}\rangle_\mathrm{eq} =-\hbar\left(\frac{1}{2}%
+f(\omega_{n,m}^{\lambda})\right)
(|v_{nm}^{\lambda(+)}|^{2}-|v_{nm}^{\lambda(-)}|^{2}).  \label{jnml}
\end{align}
To derive this, we use the relation $\langle\hat{a}_{n,m}^{\lambda\dagger} 
\hat{a}_{n^{\prime},m^{\prime}}^{\lambda^{\prime}}\rangle_\mathrm{eq}
=\delta_{n,n^{\prime}}\delta_{m,m^{\prime}}\delta_{\lambda,\lambda^{
\prime}}f(\omega_{n,m}^{\lambda})$, where $f(\omega_{n,m}^{\lambda})$ is the
Bose--Einstein distribution function with energy $\omega_{n,m}^{\lambda}$.
This result is similar to those presented in Ref.~[\citen{Zhang2014}] for
two-dimensional rotational phonons; however, Eq.~(\ref{jnml}) explicitly
reveals the dependence on CM ($n$) and branch ($\lambda$ and $m
$) of truly chiral phonons propagating along the helical axis.

\begin{figure}[ptb]
\begin{center}
\includegraphics[width=8.5cm]{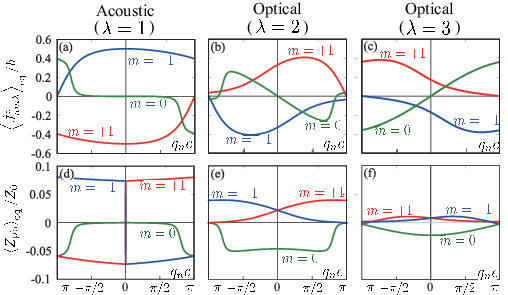} 
\end{center}
\caption{(Color online).(a) and (d) respectively show CM dependence of the
MAM and Zilch carried by phonos with different CAM for the acoustic branch ($%
\protect\lambda=1$), (b) and (e) for the first optical branch ($\protect%
\lambda=2$), and (c) and (f) for the second optical branch ($\protect\lambda=3$). 
$Z_{\text{ph}}$ is scaled by $Z_{0}\equiv 3\hbar/(cM \omega_{0})$ , with $\hbar\omega_{0}$ chosen as 1 meV.}
\label{AZ}
\end{figure}

Following the definition of Zilch for circular polarized light,\cite%
{Lipkin1964} we define the Zilch of the chiral phonon as $Z_\mathrm{ph}
\equiv  \sum_{l} \bm{u}_{l} \cdot \nabla \times \bm{u}_{l} = i \sum_{l} %
\left[  (u^{(+)}_{l})^{\dagger} \frac{\partial u^{(+)}_{l}}{\partial z} - 
(u^{(-)}_{l})^{\dagger} \frac{\partial u^{(-)}_{l}}{\partial z} \right]$,
where we assume that $\bm{u}_{l}$ depends only on $z$. Phonon chirality may
also be measured using the inner product of the MAM and wave vector.
However, the phonon Zilch involves the spatial derivative of the
wavefunction, making it an indicator that reflects the non-local spatial
structure of phonon wave functions. The derivative of $u^{(s)}_{l}$ is
discretized as $\frac{\partial}{\partial z}u^{(s)}_{l}(t)%
\equiv(u^{(s)}_{l+1}-u^{(s)}_{l})/\frac{c}{3}$. Then we have 
\begin{align}
\langle Z_\mathrm{ph}\rangle_\mathrm{eq} & = \frac{3}{c}\sum_{n}\sum_{m}
\sum_{\lambda}\frac{\hbar}{M\omega_{n,m}^{\lambda}} \left(\frac{1}{2}
+f(\omega_{n,m}^{\lambda})\right)  \notag \\
& \times \sum_{s=\pm} m_s \sin\left[q_{n}\frac{c}{3}+\alpha(m-m_s)\right]
\lvert v_{nm}^{\lambda(s)}\rvert^{2}.  \label{zilch}
\end{align}
In Fig.~\ref{AZ}, we show the CM dependence of the MAM and
Zilch at zero temperature. 
As already discussed in Ref.~[\citen{Kato2023}], the MAM continuously
changes as functions of $q_n$. On the other hand, Zilch exhibits
discontinuous jumps at $q_{n}=0$, in the acoustic mode, $\lambda=1$. The
reason for this is as follows: near $q_{n}=0$, the frequency $\omega$ of the
acoustic mode is proportional to $|q_{n}|$. For $m=\pm1$,  one of the two $%
\sin$ terms in Eq.~(\ref{zilch}) has no dependence on $m$, and gives rise to contribution proportional to $q_{n}/|q_{n}|$. 
This accounts for the
discontinuous jumps in Zilch. The other term disappears because the
corresponding eigenvector components vanish. 
The conservation law of optical Zilch originates from the duality of the electromagnetic field. However, in the case of elastic fields, no corresponding duality exists. While phonon Zilch serves as a quantitative measure of chirality, it is unlikely to be a conserved quantity. Nevertheless, as we see form Fig.~\ref{AZ}, there is a unique correspondence between the conserved CAM and the non-conserved MAM or Zilch, respectively. This result raises an interesting question regarding how MAM and Zilch are transferred inside the system under the control of the conservation of CAM.

To summarize, we constructed the wavefunction of chiral phonons in the
chiral helical crystal, and using this we formulated the electron-chiral
phonon coupling that reflects conservation of both the CM, $q_{n}$, and the CAM, $m$. We also discussed the MAM and Zilch,
and their interrelation. We expect that the framework presented in this
letter, combined with realistic electronic band structures including
spin-orbit coupling and first-principles phonon calculations, will elucidate
quantum processes of angular momentum transfer mediated by phonons in chiral
crystals~\cite{Ohe2024}.

\begin{acknowledgments}
  We thank Yusuke Kato, Tomokazu Yasuike, Shiro Komata, Hiroyasu Matsuura, Takuya Satoh, and Yoshihiko Togawa for their valuable comments.
  JK acknowledges support from JSPS KAKENHI, under Grant Nos. 23K20825 and 23H00091, the grant of OML Project by the National Institutes of Natural Sciences (NINS program No. OML012301).
    AK is supported by the JSPS KAKENHI Grants No. 22H05132.
\end{acknowledgments}

\bibliographystyle{jpsj}
\bibliography{18093Ref}

\end{document}